\documentclass[runningheads]{llncs}
\usepackage{graphicx}

\usepackage{algorithmic}
\usepackage{graphicx}
\usepackage{textcomp}
\usepackage{xcolor}
\usepackage{multirow}
\usepackage[labelformat=simple]{subcaption}
\usepackage{booktabs}
\usepackage{multirow}
\usepackage[ruled,vlined]{algorithm2e}
\usepackage{pifont}
\usepackage{amsfonts}
\usepackage{amsmath}
\usepackage{amssymb}
\usepackage{dsfont}
\usepackage[flushleft]{threeparttable}
\usepackage{subcaption}
\usepackage{makecell}
\usepackage{graphicx}
\usepackage{soul}
\usepackage{diagbox}
\usepackage{tabularx}
\usepackage{hyperref}
\usepackage{xspace}
\usepackage{adjustbox}
\hypersetup{
    colorlinks=true,
    linkcolor=black,
    filecolor=black,
    urlcolor=black,
    citecolor=black,
}
\usepackage[inline]{enumitem}

\newcommand{\method}{LDA-GCL}

\newcommand{\eg}{e.g., }
\newcommand{\ie}{i.e., }

\newcommand{\aka}{\textit{a.k.a}.~}

\newcommand{\secref}[1]{Section~\ref{#1}}
\newcommand{\figref}[1]{Fig~\ref{#1}}
\newcommand{\tableref}[1]{Table~\ref{#1}} 
\newcommand{\equref}[1]{Equation~\ref{#1}} 
\newcommand{\algref}[1]{Algorithm~\ref{#1}}

\newcommand{\hide}[1]{} %

\begin{document}
\title{Adversarial Learning Data Augmentation for Graph Contrastive Learning in Recommendation}
\titlerunning{LDA-GCL}
\author{Junjie Huang\inst{1,2} 
\and Qi Cao\inst{1} 
\and Ruobing Xie\inst{3} 
\and Shaoliang Zhang \inst{3}
\and Feng Xia \inst{3}
\and Huawei Shen\inst{1,2}\thanks{indicates corresponding author.} 
\and Xueqi Cheng\inst{1,4}}
\authorrunning{ Huang et al.}
\institute{
Data Intelligence System Research Center, \\
Institute of Computing Technology, Chinese Academy of Sciences, Beijing, China
\and
University of Chinese Academy of Sciences, Beijing, China \\
\and 
WeChat, Tencent, Beijing, China\\
\and
CAS Key Laboratory of Network Data Science and Technology,\\
Institute of Computing Technology, Chinese Academy of Sciences, Beijing, China \\
\email{\{huangjunjie17s, caoqi, shenhuawei, cxq\}@ict.ac.cn,\\
\{ruobingxie, modriczhang, xiafengxia\}@tencent.com}
}

\maketitle              %
\begin{abstract}
Recently, Graph Neural Networks (GNNs) achieve remarkable success in Recommendation.
To reduce the influence of data sparsity, Graph Contrastive Learning (GCL) is adopted in GNN-based CF methods for enhancing performance. 
Most GCL methods consist of data augmentation and contrastive loss (\eg InfoNCE).
GCL methods construct the contrastive pairs by hand-crafted graph augmentations and maximize the agreement between different views of the same node compared to that of other nodes, which is known as the \textbf{InfoMax} principle.
However, improper data augmentation will hinder the performance of GCL.
\textbf{InfoMin} principle, that the good set of views shares minimal information and gives guidelines to design better data augmentation.
In this paper, we first propose a new data augmentation (\ie edge-operating including edge-adding and edge-dropping).
Then, guided by \textbf{InfoMin} principle, we propose a novel theoretical guiding contrastive learning framework, named \underline{L}earnable \underline{D}ata \underline{A}ugmentation for \underline{G}raph \underline{C}ontrastive \underline{L}earning (\method).
Our methods include data augmentation learning and graph contrastive learning, which follow the \textbf{InfoMin} and \textbf{InfoMax} principles, respectively.
In implementation, our methods optimize the adversarial loss function to learn data augmentation and effective representations of users and items.
Extensive experiments on four public benchmark datasets demonstrate the effectiveness of \method.

\end{abstract}

\section{Introduction}
Collaborative Filtering (CF)~\cite{sarwar2001item} is to produce effective recommendations from implicit feedback (\eg clicking, rating, buying, and so on).
The interaction data can be viewed as a user-item bipartite graph.
Based on modeling such bipartite graphs, Graph Neural Networks (GNNs) ~\cite{he2017neural,wang2020disentangled,he2020lightgcn} can learn the effective node representations of users and items for personalized recommendations.
GNN model effectively utilizes the high-order graph structure information through the message-passing scheme and has achieved the-state-of-the-art results.

Although GNN models have achieved remarkable success, they still suffer from data sparsity issues.
To overcome the difficulties, Graph Contrastive Learning (GCL) in a self-supervised manner, is introduced to improve the recommendation performance.
Since no labeled data is required, GCL is considered a good solution for data sparsity issues in recommender systems~\cite{wu2021self-sigir,lin2022improving}.
GCL has two important components: data augmentation and contrastive loss.
For data augmentation, the past GCL approaches in recommendation~\cite{wu2021self-sigir} generate handcrafted graph augmentations by edge-dropping.
After data augmentation, GCL uses graph neural network models to get the node representations over multiple views.
The contrastive loss (\eg InfoNCE) leverages the mutual information maximization principle  (\textbf{InfoMax}) that aims to \textit{maximize} the correspondence between the representations of the nodes in its different augmented graphs.
However, improper data augmentation can hinder the performance of contrastive learning~\cite{yu2022graph}.
How to find the proper data augmentation is a promising research problem. 
Tian et al.~\cite{tian2020makes} investigate the research problem of what makes for good views in contrastive learning (CL).
Inspired by Information Bottleneck (IB)~\cite{tishby2000information}, they proposed the \textbf{InfoMin} principle that the good set of views shares the \textit{minimal} information necessary to perform well at the downstream task.
They find that stronger data augmentation indeed leads to decreasing mutual information and improves downstream tasks.
\textbf{InfoMin} principle offers the guideline for us to find optimal data augmentation for GCL in the recommendation.

In this paper, we propose a new graph data augmentation in recommender systems (\ie edge-operating including edge-adding and edge-dropping). 
First, to avoid randomly adding noisy edges that harm the graph structural information, we use a pre-trained model to predict the possible edges (\aka link between user $u$ and the item that is most likely to interact).
Second, we constitute the original edges and the added edges as edge candidates. 
Our augmentation strategy can better explore the diversity of graph structure than traditional edge dropping.
Third, we propose to a new adversarial \underline{L}earnable \underline{D}ata \underline{A}ugmentation for \underline{G}raph \underline{C}ontrastive \underline{L}earning (\method) framework in recommendation.
\method~ uses an edge operator model to learn the augmented graph from the candidate edges instead of randomly sampling.
Our learning data augmentation (LDA) process follows the \textbf{InfoMin} principle, while GCL framework follows the \textbf{InfoMax} principle.
\method~ optimizes an adversarial loss function to get effective embeddings for recommendation tasks.
Compared with heuristic design, our approach can automatically generate efficient graph data augmentation with \textbf{InfoMin} principle.
We analyze the effectiveness of our approach on several public benchmark datasets. 
To the best of our knowledge, it’s the first time to introduce \textbf{InfoMin} principle into GCL in recommendation.

The major contributions of this paper are summarized as follows: 
\begin{itemize}[leftmargin=*]
    \item We propose a new data augmentation method (\ie edge-operating) for contrastive learning, which includes edge-adding from a pre-trained GNN model and automatic edge-dropping from the edge candidates.
    \item Based on \textbf{InfoMin} and \textbf{InfoMax} principles, we proposed a new adversarial framework for learning efficient data augmentation, called \method.
    \method~consists of learning data augmentation and graph contrastive learning.
    \item We conduct extensive experiments on four real-world public benchmark datasets. Experimental results demonstrate the effectiveness of the proposed \method.
\end{itemize}

\section{Preliminary}
\label{sec:preliminary}

\subsection{Bipartite Graph in Recommendation}

As the fundamental recommender system, collaborative filtering (CF) can be modelled as a user-item bipratite graph as $G = (\mathcal{U}, \mathcal{I}, \mathcal{E})$, where $\mathcal{U}$ is the user set, $\mathcal{I}$ is the item set and $\mathcal{E} \subseteq \mathcal{U} \times \mathcal{I}$ is the  inter-set edges. 
$\mathcal{E}$ can be denoted as the user-item interaction matrix $\mathbf{R}\in \{0, 1\}^{|\mathcal{U}| \times |\mathcal{I}|}$.
The adjacency matrix
$
\mathbf{A}=\left[\begin{array}{cc}
\mathbf{0} & \mathbf{R} \\
\mathbf{R}^{\top} & \mathbf{0}
\end{array}\right]
$
is also widely used in~\cite{he2020lightgcn}.

\subsection{GNN-based Collaborative Filtering}
\label{sec:gnn_cf_pre}
Based on the bipartite graph definition, the general GNN-based collaborative filtering methods follow the message-passing scheme to generate informative representations for users and items:
\begin{equation}
z_{w}^{l} =f_{\text {aggregate }}\left(\left\{z_{v}^{l-1} \mid v \in \mathcal{N}_{w} \cup\{w\}\right\}\right), 
z_{w} =f_{\text {update }}\left(\left[z_{w}^{0}, z_{w}^{1}, \ldots, z_{w}^{L}\right]\right),
\end{equation}
where $\mathcal{N}$ denotes the neighbor set of node $w$ in bipartite graph $G$ and $L$ denotes the number of GNN layers. $z^{0}$ is the learnable initial embeddings.
$f_{\text {aggregate }}$ and $f_{\text {update }}$ are aggregate function and update function designed by different models.
Specifically, the state-of-the-art method (\ie LightGCN~\cite{he2020lightgcn}) removes nonlinear activation and feature transformation of NGCF~\cite{wang2019neural} and applies a simple weighted sum aggregator:
\begin{equation}
    \label{eq:lightgcn}
    Z^{l+1} =\left(\mathbf{D}^{-\frac{1}{2}} \mathbf{A} \mathbf{D}^{-\frac{1}{2}}\right) Z^{l}, 
    Z = \frac{1}{L+1}(Z^0 + Z^1 + \cdots + Z^L),
\end{equation}
where $\mathbf{D}_{ii}=\sum_j \mathbf{A}_{ij}$ is the diagonal matrix and $Z^0$ is initial trainable embeddings.
After obtaining the final embedding $Z$, the inner product is used to predict how likely user $u$ would adopt item $i$ by $\hat{y}_{ui}=z_u ^T z_i$.
Most GNN-based CF methods (\eg NGCF~\cite{wang2019neural}, DGCF~\cite{wang2020disentangled}, and LightGCN~\cite{he2020lightgcn}) use the pairwise Bayesian Personalized Ranking (BPR) loss function for the model training:
\begin{equation}
\mathcal{L}_{\text{BPR}}=\sum_{(u, i, j) \in \mathcal{O}}-\log \sigma\left(\hat{y}_{u i}-\hat{y}_{u j}\right),
\end{equation}
where $\mathcal{O} = \{(u, i, j) | (u, i)\in \mathcal{O}^+, (u, j)\in \mathcal{O}^-\}$, $\mathcal{O}^+$ and $\mathcal{O}^-$ are the observed and unobserved interactions, respectively.

\subsection{Graph Contrastive Learning in Recommendation}

To overcome the data sparsity issues, Graph Contrastive Learning (GCL) is introduced into recommender systems.
GCL first applies data augmentation and then contrasts the two augmented samples. 
Common data augmentation is the perturbation of the graph structure due to the absence of node features.
Specifically, SGL~\cite{wu2021self-sigir} proposes edge-dropping, node-dropping, and random walk data augmentation strategies. 
After data augmentation, the augmented views of the same user node are treated as the positive pairs (\ie $\{(z_u^{\prime}, z_u^{\prime \prime})\}$), and the views of different user nodes are treated as the negative pairs (\ie $\{(z_u^{\prime}, z_v^{\prime \prime}\}$)). 
Following SimCLR~\cite{chen2020simple}, the contrastive loss (\ie InfoNCE~\cite{gutmann2010noise}) is adopted to maximize the agreement of positive pairs and minimize that of negative pairs by
\begin{equation}
\label{eq:info_nce}
\mathcal{L}^{\mathcal{U}}_{\text{NCE}}=\sum_{u \in \mathcal{U}}-\log \frac{\exp \left(sim\left(\mathbf{z}_{u}^{\prime}, \mathbf{z}_{u}^{\prime \prime}\right) / \tau\right)}{\sum_{v \in \mathcal{U}} \exp \left(sim\left(\mathbf{z}_{u}^{\prime}, \mathbf{z}_{v}^{\prime \prime}\right) / \tau\right)}, 
\end{equation}
where $\tau$ is the temperature hyper-parameters and $sim$ is the similarity function (\eg cosine function).
Analogously, contrastive loss is also adopted on the item side (\ie $\mathcal{L}^{\mathcal{I}}_{\text{NCE}}$).
The final contrastive loss is the combination of two losses as $\mathcal{L}_{\text{NCE}} = \mathcal{L}^{\mathcal{U}}_{\text{NCE}} + \mathcal{L}^{\mathcal{I}}_{\text{NCE}} $.
It is worthwhile mentioning that the contrastive learning~\cite{wu2021self-sigir,lin2022improving} in recommender systems usually adopts the joint learning strategy to train their model instead of pre-training and fine-tuning strategies~\cite{jin2020self}. 
In other words, both pretext tasks and downstream tasks are optimized jointly~\cite{wu2021self}.
Wu et al.~\cite{wu2021self-sigir} demonstrate that joint training will achieve better performance, the pretext tasks and downstream tasks are mutually enhanced with each other.

\section{Methodology}
\label{sec:model}
\begin{figure*}[htp]
    \centering
    \includegraphics[width=\textwidth]{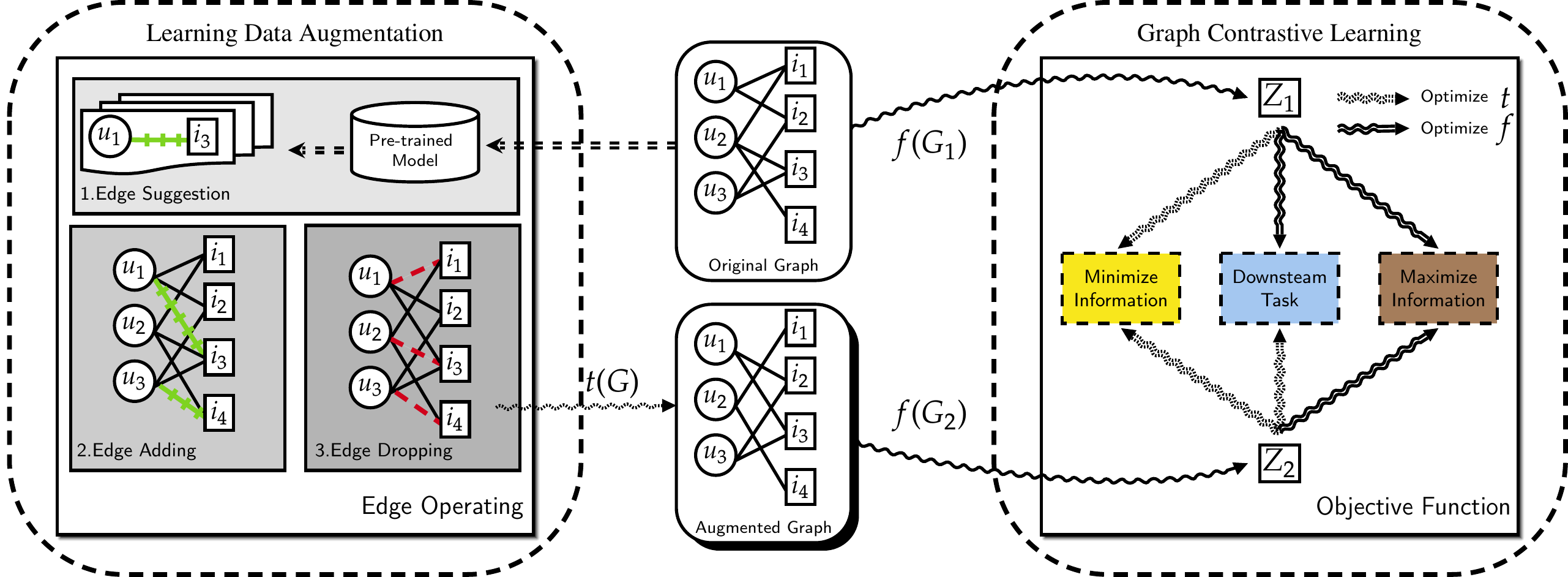}
    \caption{Illustration of our framework \method. \method~ includes learning data augmentation and graph contrastive learning. }
    \label{fig:framework}
\end{figure*}

In this section, we introduce our framework for learning data augmentation for recommender systems.
In \figref{fig:framework}, our methods include pre-trained edge candidate generation, edge operating, and min-max objective functions.

\subsection{Graph Data Augmentation With Edge Operating}
\label{sec:data-aug}

As mentioned in \secref{sec:preliminary}, the existing data augmentation in recommender systems is generally edge-dropping.
The other node dropping and random walk dropping can also be regarded as different strategies of edge-dropping.
Edge-dropping can be elaborated as follows:
\begin{equation}
s_{1}(G) = \mathbf{A}_1 = \mathbf{A} \odot \mathbf{M}_1, \quad s_{2}(G) = \mathbf{A}_2 = \mathbf{A} \odot \mathbf{M}_{2},
\end{equation}
where $\odot$ is the Hadamard product and $\mathbf{M}_1, \mathbf{M}_2 \in \{0, 1\}^{|V| \times |V|}$ are two masking matrices to be applied on the original graph $G$ to generate two augmented graph adjacency matrix $\mathbf{A}_1$ and $\mathbf{A}_2$.
The node dropping and edge random walk strategies just use different masking matrices (\eg the row or column sum of a masking matrix with node dropping will be zero).
In practice, sampling edges follow a uniform distribution to keep $(1-\rho)\times |\mathcal{E}|$ edges, where $\rho$ is the edge-dropping ratio. $\rho$ is usually set to a small value (\eg 0.1).

However, the only edge-dropping strategies will suffer the data sparsity issues~\cite{wu2021self-sigir}.
We propose a new data augmentation in recommender systems (\ie \textbf{edge-operating} including both edge-adding and edge-dropping).
Compared with the edge perturbation in self-supervised on graphs~\cite{wu2021self}, edge-operating in recommendation systems faces some challenges.
First, the complexity of randomly sampling edges from $\mathbf{A}$ is $\mathcal{O}((|V|)^2)$ and it is not acceptable in large-scale recommender systems,  which have million-level users and items.
Second, randomly adding edges to $\mathbf{A}$ will introduce noises. 
Therefore, we propose to first construct edge candidates and then sample from these edge candidates.
As shown in \figref{fig:framework}, we firstly train a GNN model (\eg LightGCN) to predict the item preference of user $u$ (\ie Edge Suggestion).
We choose the top-$K_u$ items for our candidate, where $K_u$ is the degree of the user $u$ in the user-item interaction matrix $\mathbf{R}$. 
After adding edges, our final edge candidates consist of original edges $\mathcal{E}_0$ and suggested edges $\mathcal{E}_1$ ($|\mathcal{E}_0| = |\mathcal{E}_1|)$.
We can sample edges from edge candidates and conduct graph contrastive learning (see \secref{sec:ablation_study}).

\subsection{Learning Data Augmentation}
After introducing edge-operating augmentation, we further propose to use a learnable edge operator model $t$ to generate informative data augmentation instead of random sampling.
In order to learn the graph data augmentation, we use a Multi-layer Perception (MLP) to learn the weight for every edge candidate $e_{u,i}$ as follows:
\begin{equation}
    \omega_{u,i}=\operatorname{MLP}\left(\left[z_{u} \odot z_{i}\right] \|  \mathds{1}_{\mathcal{E}}(e_{u,i}) \right),
\end{equation}
where $\odot$ is the Hadamard product, $z_u$ and $z_i$ are the embeddings for user $u$ and item $i$, $\|$ is the concatenation operator and $\mathds{1}_{\mathcal{E}}(e_{u,i})$ indicates if edge $e_{u,i}$ belongs to original or added edges.
Then, we use the  Gumbel-Max reparameterization~\cite{DBLP:conf/iclr/JangGP17} to get the probability $p_{u,i}$ for edge $e_{u,i}$ by
\begin{equation}
    \label{eq:p_ui}
    p_{u,i} = \mathrm{sigmoid}(\frac{(\log\delta - \log(1-\delta) + \omega_{u,i})}{\tau}),
\end{equation}
where $\delta \sim$ Uniform(0,1) and $\tau$ is the temperature hyperparameter. 
This style of edge learning has also been used in parameterized explanations and adversarial attacks of  GNNs~\cite{luo2020parameterized,tao2021single}.
Further, we use $p_{u,i}$ to construct augmented graphs
$
t(G)= \mathbf{A}^{\prime} = \left(\begin{array}{cc} \mathbf{0} & \mathbf{P} \\ \mathbf{P}^{\top} & \mathbf{0}\end{array}\right),
$
where $\mathbf{P}\in R^{|\mathcal{U}|\times |\mathcal{I}|}$ is the probability matrix from \equref{eq:p_ui}.
We can apply GNN model (\ie LightGCN in this paper) to original graph $G_1=G$ with the adjacency matrix $\mathbf{A}$ and augmented graph $G_2=t(G)$ with the adjacency matrix $\mathbf{A^{\prime}}$, and  get the embeddings $Z_1$ and $Z_2$ by \equref{eq:lightgcn}.

\subsection{Objective Function} 
\label{sec:loss_function}
Inspired by related works on graph contrastive learning ~\cite{suresh2021adversarial,xu2021infogcl}, we use an adversarial loss function to find good graph augmentations to enhance GCL in recommendation. 
The objective functions are defined as follows:
\begin{equation}
\label{eq:loss_function}
\begin{aligned}
\min_{t}\  \lambda_t I(f(G); f(t(G)))  + \mathcal{L}(f(t(G)), y) \\
\max_{f}\  I(f(G); f(t(G))) - \mathcal{L}(f(G), y),
\end{aligned}
\end{equation}
where $I(X_1; X_2)$ is the mutual information between two random variables $X_1$ and $X_2$, $t$ is the data augmentation learner, $f$ is the GNN encoder (\ie LightGCN in this paper) and $\mathcal{L}$ is the task relevant supervised loss function (\ie BPR loss function in this paper).
$\lambda_t$ is used to control the influence of $I$ for $t$.

In order to estimate the Mutual Information (MI), we choose InfoNCE Estimator~\cite{chen2020simple,gutmann2010noise}, which is one of the most popular lower-bound to the mutual information.
Based on InfoNCE Estimator in \equref{eq:info_nce}, we can map $ I(f(G), f(t(G))$ in \equref{eq:loss_function} to:
\begin{small}
\begin{equation}
    I(f(G), f(t(G)) \rightarrow -\mathcal{L}_{\text{NCE}}=\frac{1}{B} \sum_{i=1}^{B} \log \frac{\exp \left(sim\left(z_{i, 1}, z_{i, 2}\right)\right)}{\sum_{i^{\prime}=1, i^{\prime} \neq i}^{B} \exp \left(sim\left(z_{i, 1}, z_{i^{\prime}, 2}\right)\right)},
\end{equation}
\end{small}

where $sim$ is the cosine similarity to measure the agreement between two representations, $z$ is the node representaton encoded by $f(G)$ and $f(t(G))$, and $B$ is the batch size.
\equref{eq:loss_function} is the min-max optimize problem, we use the iterative training approach used in adversial traning~\cite{adversial_training}.

When we fix $t$, we have following loss function:
\begin{equation}
    \label{eq:fix_t}
   \mathcal{L}_{f} =\mathcal{L}_{\text{BPR}}(f(G), y) +  \lambda_{ssl}\mathcal{L}_{\text{NCE}}\ (f(G), f(t(G))) +  \lambda_{reg} \|f\|^2_{2},
\end{equation}
where $\lambda_{ssl}$ and $\lambda_{reg}$ are the hyper-parameters to control the weights of the InfoNCE loss function and the regularization term.
\equref{eq:fix_t} is also used in self-supervised graph learning for recommendation~\cite{wu2021self-sigir}.
We follow the setting of $\lambda_{ssl}=0.1$ in ~\cite{wu2021self-sigir}.
\equref{eq:fix_t} leverages the mutual information maximization principle (\textbf{InfoMax}) to capture as much information as possible about the stimulus. When we fix $f$, we have following loss function:
\begin{equation}
   \label{eq:fix_f}
   \mathcal{L}_{t} = \mathcal{L}_{\text{BPR}}(f(t(G)), y) -\lambda_2\mathcal{L}_{\text{NCE}}\ (f(G), f(t(G))) + \lambda_{reg} \|t\|^2_{2},
\end{equation}
where $\lambda_2 = \lambda_t \times \lambda_{ssl}$ and $\lambda_{reg}$ are the hyper-parameters to control the weights of the InfoNCE loss function and the regularization term.
Contrary to \equref{eq:fix_t}, \equref{eq:fix_f} follow the \textbf{InfoMin} principle to find augmented graph that share the minimal information necessary to perform well at the downstream task~\cite{tian2020makes}.

\subsection{Training \method}

\begin{algorithm}[t]
  \caption{\method~ Training Algorithm}
  \label{alg:algorithm1}
  \begin{algorithmic}[1]
    \renewcommand{\algorithmicrequire}{\textbf{Input:}}
    \renewcommand{\algorithmicensure}{\textbf{Output:}}
    \REQUIRE {
      Original bipartite graph $G(\mathcal{U},\mathcal{I}, \mathcal{E})$; 
      Pre-trained GNN encoder $f_0$;
      GNN encoder $f$;
      Edge operator model $t$;
      Epoch $T$;
    }
    
    \ENSURE{
      Node representation $Z$
    }
    \\
    
    \STATE{Generate added edges $\mathcal{E}_1$ from pre-trained model $f_0$.}
    \STATE{Merge added edges $\mathcal{E}_1$ and original edges $\mathcal{E}$ into edge candidates $\mathcal{E}_2$.}
    \STATE{Initialize the parameters of edge operator model $t$ and GNN encoder $f$}

    \FOR{$epoch=1,...,T$}
    \FOR{each mini-batch interactions $B=\{(u_1, i_1, i_2)\}$}
    \STATE{Get node set $V$ with user set $U$ and item set $I$ in mini-batch data}
    
    \tcc{ Optimize $t$}
    
    \STATE{Freeze GNN encoder $f$; unfreeze edge operator $t$}
    \STATE{Apply $t$ on $\mathcal{E}_2$ to get augmented graph $t(G)$ and Apply $f$ to get the embeddings $Z_1, Z_2$ for node $V$ from $G$}
    \STATE{Compute loss in \equref{eq:fix_f} with $Z_1$ and $Z_2$; Back propagation, update $t$.}
    
    \tcc{Optimize $f$}
    \STATE{Freeze edge operator $t$; unfreeze of GNN encoder $f$ }
    \STATE{Apply $t$ on $\mathcal{E}_2$ to get augmented graph $t(G)$ and Apply $f$ to get the embeddings $Z_1, Z_2$ for node $V$ from $G$}
    \STATE{Compute loss in \equref{eq:fix_t} with $Z_1$ and $Z_2$; Back propagation, update $f$.}

    \tcc{Judge early stopping condition}
    \IF{$Z_1$ match the early stopping condition}
    \STATE{Stop training algorithm; Return the best GNN encoder $f_{opt}$}
    \ENDIF
    \ENDFOR
    \ENDFOR
    \RETURN {$Z = f_{opt}(G)$}
  \end{algorithmic}
\end{algorithm}

We briefly summarize  \method~  training process in \algref{alg:algorithm1}.
As we disscussed in \secref{sec:loss_function}, \algref{alg:algorithm1} is iterative adversarial training procedure.
We use two optimizers to optimize $f$ and $t$, separately, which is similar to the training of GANs~\cite{goodfellow2014generative}.
In this paper, we use the LightGCN as the pre-trained GNN, other pre-trained models were also tried but with worse results (see \secref{sec:ablation_study}).

\section{Experiments}
\label{sec:experiments}

In this section, we conduct experiments to evaluate the effectiveness of the proposed frameworks with comparison to state-of-the-art methods. 
Specifically, we aim to answer the following research questions:
\begin{itemize}[leftmargin=*]
    \item \textbf{RQ1}: How does \method~ perform in recommendation tasks as compared with the state-of-the-art CF models and GCL models?
    \item \textbf{RQ2}: If \method~ performs well, what component benefits our \method~ in collaborative filtering tasks?     
    \item \textbf{RQ3}: What hyper-parameters affect the effectiveness of the proposed \method? 
\end{itemize}

\subsection{Experimental Settings}

\begin{table}
\vspace{-2em}
\centering
	\caption{Statistics of the datasets used in this paper.}
	\label{tb:dataset}
	\scalebox{0.8}{}{
	\begin{tabular}{c *{4}{r}}
		\toprule
		\textbf{Datasets} & \textbf{\#Users} & \textbf{\#Items} & \textbf{\#Interactions} & \textbf{\%Density}\\
		\midrule
		Yelp 			& 45,478 & 30,709 & 1,777,765 & 0.127 \\
		Gowalla         & 29,859 & 40,989 & 1,027,464 & 0.084 \\
		Amazon-Book 	& 58,145 & 58,052 & 2,517,437 & 0.075 \\
		Alibaba-iFashion & 300,000 & 81,614 & 1,607,813 & 0.007  \\
		\bottomrule
	\end{tabular}
	}
\vspace{-1em}
\end{table}
\vspace{-1em}
\subsubsection{Datasets}
We conduct our experiments on four  public benchmark datasets, which are widely used for recommender systems: Yelp~\cite{chen2019pog}, Gowalla~\cite{cho2011friendship}, Amazon-Book~\cite{mcauley2015image} and Alibaba-iFashion~\cite{chen2019pog}.
For the Yelp and Amazon Books datasets, we filter out users and items with fewer than 15 interactions.
The Alibaba-iFashion is a large and sparse dataset.
The statistics of the datasets used in this paper are summarized in \tableref{tb:dataset}. 
Our experimental settings are close to \cite{lin2022improving}, but we remove the duplicated user-item interactions on the Yelp dataset.
For each dataset, we randomly select 80\% of interactions as training data and 10\% of interactions as validation data. 
The remaining 10\% interactions are used for testing model performance. 
We run such training/validation/testing data split 5 times to report the average scores.
We uniformly sample one negative item for each positive instance to form the training set.

\subsubsection{Baselines}
We compare the proposed method with the following different kinds of baseline methods including Matrix Factorization (MF) methods, Graph Neural Networks (GNN) methods, and Graph Contrastive Learning (GCL) methods:
\begin{itemize}
    \item \textbf{BPRMF}~\cite{rendle2012bpr}: It optimizes the BPR loss function (see \secref{sec:gnn_cf_pre}) to learn the user and item  representations with matrix factorization (MF) framework.
    \item \textbf{NeuMF}~\cite{he2017neural}: It is a generic matrix factorization model using a multilayer perceptron (MLP) to learn the user-item interaction function. 
    NeuMF uses the pointwise binary cross-entropy loss function to optimize the model.
    \item \textbf{DMF}~\cite{xue2017deep}: It is a matrix factorization model using a deep learning architecture to learn the representations of users and items.
    It uses a normalized cross entropy loss to optimize the DMF model.
    \item \textbf{NGCF}~\cite{wang2019neural}: NGCF integrates the bipartite graph structure into the embedding learning process.
    It uses standard GCN~\cite{kipf2017semi} to enhance CF methods.
    \item \textbf{DGCF}~\cite{wang2020disentangled}: DGCF is a GNN model to disentangle user intents  factors and yield disentangled representations for user and item.
    \item \textbf{LightGCN}~\cite{he2020lightgcn}: It simplifies the design of NGCF and devises a light graph convolution for training efficiency and generation ability.
    It can be viewed as the state-of-the-art GNN-based method.
    \item \textbf{SGL}~\cite{wu2021self-sigir}: SGL introduces several data augmentations (\ie edge-dropping, node-dropping, and random-walk) and adopts GCL to enhance recommendation. 
    In this paper, we adopt the most powerful SGL-ED as the instantiation of SGL.
     \item \textbf{SimGCL}~\cite{yu2022graph}: It  is a simple GCL method  with uniform noises to the embedding space.
    \item \textbf{NCL}~\cite{lin2022improving}: It explicitly incorporates the structural neighbors and semantic neighbors into contrastive pairs. 
    It can be viewed as the most advanced GCL in recommendation.
\end{itemize}
\subsubsection{Metrics}
To evaluate the performance of top-$N$ recommendation, we adopt two widely used metrics Recall@$N$ and NDCG@$N$ , where $N$ is set to 10, 20 and 50 for consistency. 
Following ~\cite{he2020lightgcn}, we adopt the full-ranking protocol~\cite{wu2021self-sigir}, which ranks all the candidate items that the user has not interacted with.

\subsection{Implementation Details}
We implement the proposed model and all the baselines based on RecBole\footnote{https://recbole.io/}~\cite{zhao2021recbole}, which is developed based on PyTorch for reproducing and developing recommendation algorithms in a unified, comprehensive, and efficient framework for research purpose.
To be a fair comparison, we use Adam optimizer to optimize all the models.
The embedding size is set to 64. 
The batch size is set to 4,096 and all the parameters are initialized by the default Xavier distribution. 
For NCL, we use the authors' released code from github\footnote{https://github.com/RUCAIBox/NCL}. 
We follow the authors' suggested hyper-parameter settings. 
We adopt early stopping with the patience of 10 epochs to prevent overfitting, and NDCG@10 is set as the early stopping indicator. 
All experiments run on an NVIDIA Tesla V100S GPU (32GB) and 
Intel(R) Xeon(R) Silver 4310 CPUs (250GB).

\subsection{Performance Comparision (RQ1)}

\tableref{tab:exp-main} shows the performance comparison of the proposed \method~ and other baseline methods on four datasets.
From \tableref{tab:exp-main}, we can find that:
\begin{itemize}[leftmargin=*]
    \item For MF-based methods, BPRMF outperforms NeuMF and DMF on all datasets (\eg 17.9\% in Recall@10 and 22.4\% in NDCG@10 on Gowalla, compared with NeuMF).
    It can be due to the reason that it is difficult for MLP to learn the dot product~\cite{rendle2020neural} and the pointwise loss function (\eg Binary Cross-Entropy loss function in NeuMF) is less effective than the pairwise loss function (\eg BPR loss function in BPRMF).
    \item Compared to MF-based methods (\eg BPRMF), GNN-based methods exhibit better performance on most datasets.
    GNN-based methods utilize the structural information of bipartite graphs into representations by introducing the powerful GNNs. 
    Among all the GNN-based models, LightGCN performs best in most datasets, demonstrating the effectiveness of simplified architecture~\cite{he2020lightgcn}.
    BPRMF performs even better than NGCF on Gowalla and Amazon-Book.
    Similar results are also reported in ~\cite{lin2022improving}, again showing that the heavy GNN architecture will overfit and limit the performance.
    For the disentangled representation learning method DGCF, we limit the number of layers to 2 because of GPU memory limitations on the Amazon-Book dataset.
    DGCF is worse than LightGCN, especially on the sparse dataset (\eg 11.2\% in Recall@10 and 11.5\% in NDCG@10 on Alibaba-iFashion).
    The low number of factors for disentanglement (default value is 4) may be the limit of DGCF.
    \item For the GCL-based baseline methods (\ie SGL, SimGCL, and NCL), GCL-based methods consistently outperform other supervised GNN-based methods on all datasets, which shows the effectiveness of GCL for improving performance. 
    With uniform noises, SimGCL outperforms SGL on most datasets.
    However, SimGCL performs poorly on some datasets (\eg Alibaba-iFashion). 
    The hyper-parameters in SGL and SimGCL (\eg data augmentation strategy, edge dropout ratio, and contrastive ratio) will severely impact the performance.
    Compared with SGL and SimGCL, NCL achieves significant improvements on most datasets, which is consistent with the results reported in~\cite{lin2022improving}.
    We consider NCL as the most competitive baseline.
    NCL incorporates the neighborhood-enriched contrastive learning objectives and achieves better results than SGL and SimGCL on most datasets.
    \item For our \method, we can find that \method~ outperforms all baselines and significantly performs better than the state-of-the-art NCL on most datasets. 
    It is worth mentioning that our method achieves the largest improvement on the largest and sparsest dataset (\ie Alibaba-iFashion). 
    On Alibaba-iFashion, \method~outperforms NCL 23.5\% and 25.28\%  in Recall@10 and NDCG@10, respectively.
    And it also surpasses the SGL on the Alibaba-iFashion dataset by about 10\%.
    The experimental results show the effectiveness of the method proposed in this paper.
\end{itemize}

\begin{table*}[ht]
\hspace{-2em}
\scalebox{0.8}{
\begin{threeparttable}
\centering
\caption{Performance Comparison of Different Baseline Models}
\label{tab:exp-main}

\begin{tabular}{@{}c|c|ccc|ccc|cccc}
\toprule
 & &\multicolumn{3}{|c|}{Matrix Factorization} & \multicolumn{3}{|c|}{Graph Neural Networks} & \multicolumn{4}{|c}{Graph Contrastive Learning}\\
\midrule
Dataset  & Metric & BPRMF & NeuMF & DMF  & NGCF & DGCF & LightGCN & SGL & SimGCL & NCL & \method~ \\ 

\midrule

\multirow{6}{*}{Yelp} 
& Recall@10 & 0.0499 & 0.0367 & 0.0372 & 0.0514 & 0.0606 & 0.0616 & 0.0664  & \underline{0.0743} & 0.0713 &\textbf{0.0751}$^*$ \\ 
& Recall@20 & 0.0829 & 0.0629 & 0.0631 & 0.0857 & 0.0987 & 0.1001 & 0.1072  & \underline{0.1185} & 0.1135 & \textbf{0.1190}$^*$  \\ 
& Recall@50 & 0.1549 & 0.1227 & 0.1215 & 0.1596 & 0.1798 & 0.1817 & 0.1928  & \underline{0.2068} & 0.1997 & \textbf{0.2101}$^*$  \\
& NDCG@10 & 0.0335 & 0.0242 & 0.0248 & 0.0346 & 0.0412 & 0.0419 & 0.0456  & \underline{0.0515} & 0.0489 & \textbf{0.0518}$^*$ \\
& NDCG@20 & 0.0438 & 0.0324 & 0.0327 & 0.0453 & 0.0530 & 0.0538 & 0.0581 &  \underline{0.0652}  & 0.0619 & \textbf{0.0653}$^*$  \\
& NDCG@50 & 0.0622 & 0.0477 & 0.0476 & 0.0642 & 0.0738 & 0.0748 & 0.0801  & \underline{0.0878} & 0.0841 &\textbf{0.0886}$^*$ \\
\midrule

\multirow{6}{*}{Amazon-Book} 
& Recall@10 & 0.0619 & 0.0442 & 0.0313 & 0.0575 & 0.0787 & 0.0783 & 0.0844 & 0.0872 &\underline{0.0947} & \textbf{0.0975}$^*$ \\
& Recall@20 & 0.0971 & 0.0726 & 0.0522 & 0.0920 & 0.1191 & 0.1210 & 0.1281 & 0.1251 & \underline{0.1395} & \textbf{0.1456}$^*$  \\
& Recall@50 & 0.1676 & 0.1331 & 0.0984 & 0.1624 & 0.1965 & 0.2055 & 0.2117 & 0.1934 & \underline{0.2201} & \textbf{0.2346}$^*$ \\
& NDCG@10 & 0.0431 & 0.0295 & 0.0216 & 0.0400 & 0.0563 & 0.0553 & 0.0606 & 0.0643 & \underline{0.0685} & \textbf{0.0699}$^*$\\
& NDCG@20 & 0.0537 & 0.0382 & 0.0280 & 0.0505 & 0.0681 & 0.0682 & 0.0739 & 0.0758 & \underline{0.0822} & \textbf{0.0845}$^*$  \\
& NDCG@50 & 0.0721 & 0.0539 & 0.0400 & 0.0688 & 0.0887 & 0.0902 & 0.0956 & 0.0936 &\underline{0.1034} & \textbf{0.1078}$^*$\\
 \midrule

\multirow{6}{*}{Gowalla} 
& Recall@10 & 0.1040 & 0.0882 & 0.0634 & 0.0992 & 0.1343 & 0.1355 & 0.1386 & 0.1487  & \underline{0.1496} & \textbf{0.1505} \\
& Recall@20 & 0.1525 & 0.1307 & 0.0945 & 0.1462 & 0.1917 & 0.1969 & 0.1969 & 0.2123 & \underline{0.2131} & \textbf{0.2144}  \\
& Recall@50 & 0.2476 & 0.2161 & 0.1559 & 0.2383 & 0.2972 & 0.3093 & 0.3055 & 0.3208 & \underline{0.3228} & \textbf{0.3284}$^*$  \\
& NDCG@10 & 0.0738 & 0.0603 & 0.0450 & 0.0703 & 0.0963 & 0.0961 & 0.0999 & 0.1078 & \underline{0.1081} & \textbf{0.1085}  \\
& NDCG@20 & 0.0878 & 0.0727 & 0.0540 & 0.0838 & 0.1127 & 0.1136 & 0.1166 & 0.1259 & \underline{0.1263} & \textbf{0.1268}  \\
& NDCG@50 & 0.1109 & 0.0935 & 0.0692 & 0.1062 & 0.1384 & 0.1411 & 0.1431 & 0.1525 & \underline{0.1534} & \textbf{0.1547}  \\
\midrule

\multirow{6}{*}{Alibaba-iFashion} 
& Recall@10 & 0.0297 & 0.0157 & 0.0138 & 0.0355 & 0.0361 & 0.0402 & \underline{0.0518} & 0.0450 & 0.0490 & \textbf{0.0605}$^*$ \\
& Recall@20 & 0.0458 & 0.0264 & 0.0229 & 0.0565 & 0.0549 & 0.0612 & \underline{0.0774} & 0.0651 & 0.0729 & \textbf{0.0882}$^*$ \\
& Recall@50 & 0.0784 & 0.0501 & 0.0443 & 0.0994 & 0.0910 & 0.1015 & \underline{0.1258} & 0.1029 & 0.1178 & \textbf{0.1381}$^*$ \\
& NDCG@10 & 0.0158 & 0.0079 & 0.0071 & 0.0185 & 0.0194 & 0.0216 & \underline{0.0280} & 0.0252 & 0.0267 & \textbf{0.0335}$^*$  \\
& NDCG@20 & 0.0199 & 0.0106 & 0.0094 & 0.0237 & 0.0241 & 0.0269 & \underline{0.0344} & 0.0303 & 0.0328 & \textbf{0.0405}$^*$ \\
& NDCG@50 & 0.0264 & 0.0152 & 0.0137 & 0.0323 & 0.0313 & 0.0350 & \underline{0.0440} & 0.0378 & 0.0417 & \textbf{0.0504}$^*$  \\

\bottomrule
\end{tabular}

\begin{tablenotes}
      \item The best result is \textbf{bolded} and the second result is \underline{underlined}. $^*$ indicates the statistical significance for $p<$ 0.05.
\end{tablenotes}
\end{threeparttable}
}
\end{table*}
\subsection{Benefits of \method~ (RQ2)}
\label{sec:add_graph}

Based on the previous experimental results, we analyze the advantages of our model in this subsection.
We choose Gowalla and Alibaba-iFashion to analyze the reasons for the advantage in the previous subsection. 
We randomly resplit datasets as the previous subsection does.

\vspace{-1em}
\subsubsection{Sparsity Analysis}
\begin{figure}[!t]
  \centering
\begin{minipage}[t]{0.48\linewidth}
\centering
  \begin{subfigure}[t]{0.48\linewidth}
    \includegraphics[width=\linewidth]{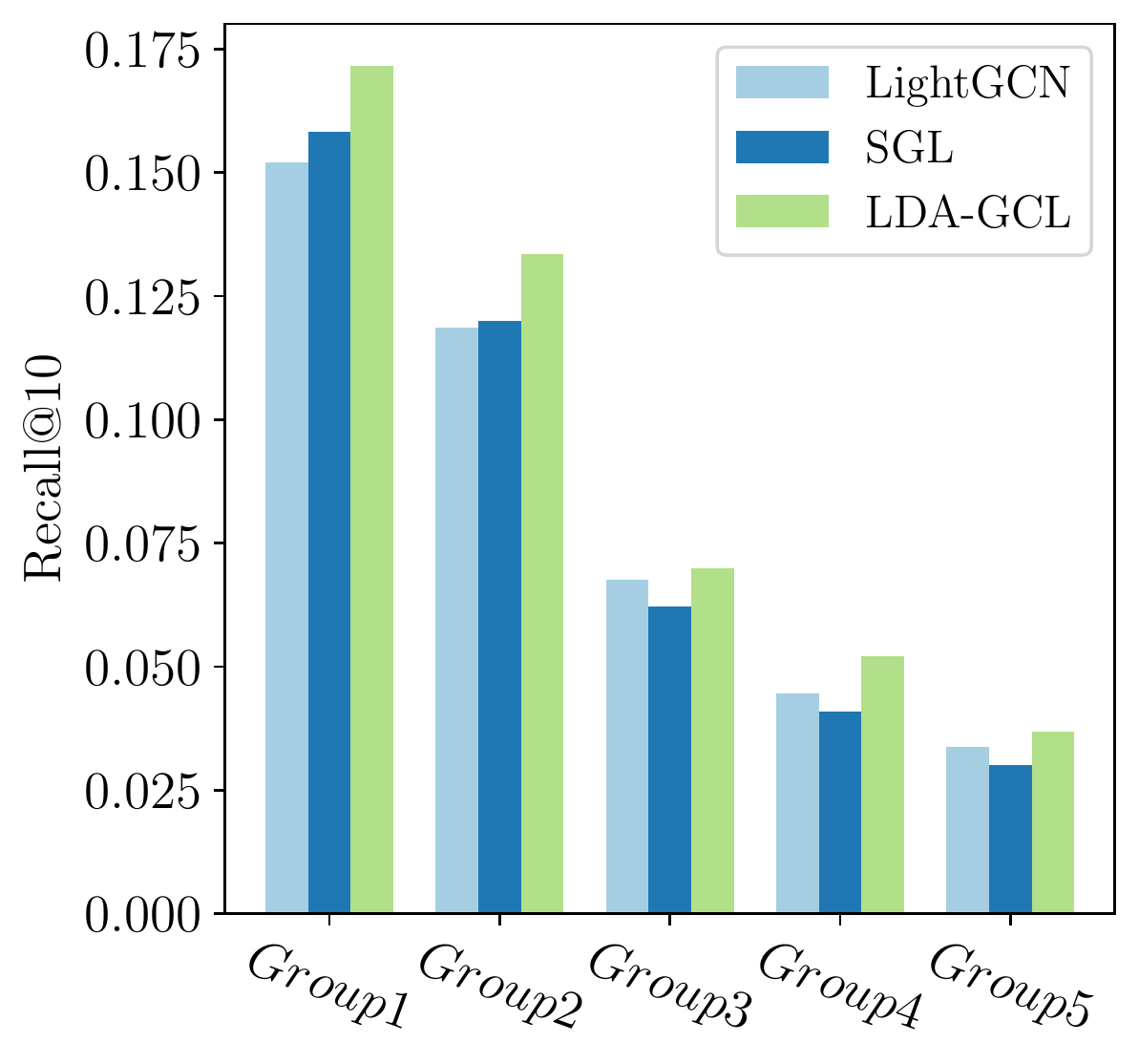}
    \caption{\small Gowalla}
    \label{fig:group-analysis-1}
  \end{subfigure}
  \hfill
  \begin{subfigure}[t]{0.48\linewidth}
  \centering
  \includegraphics[width=\linewidth]{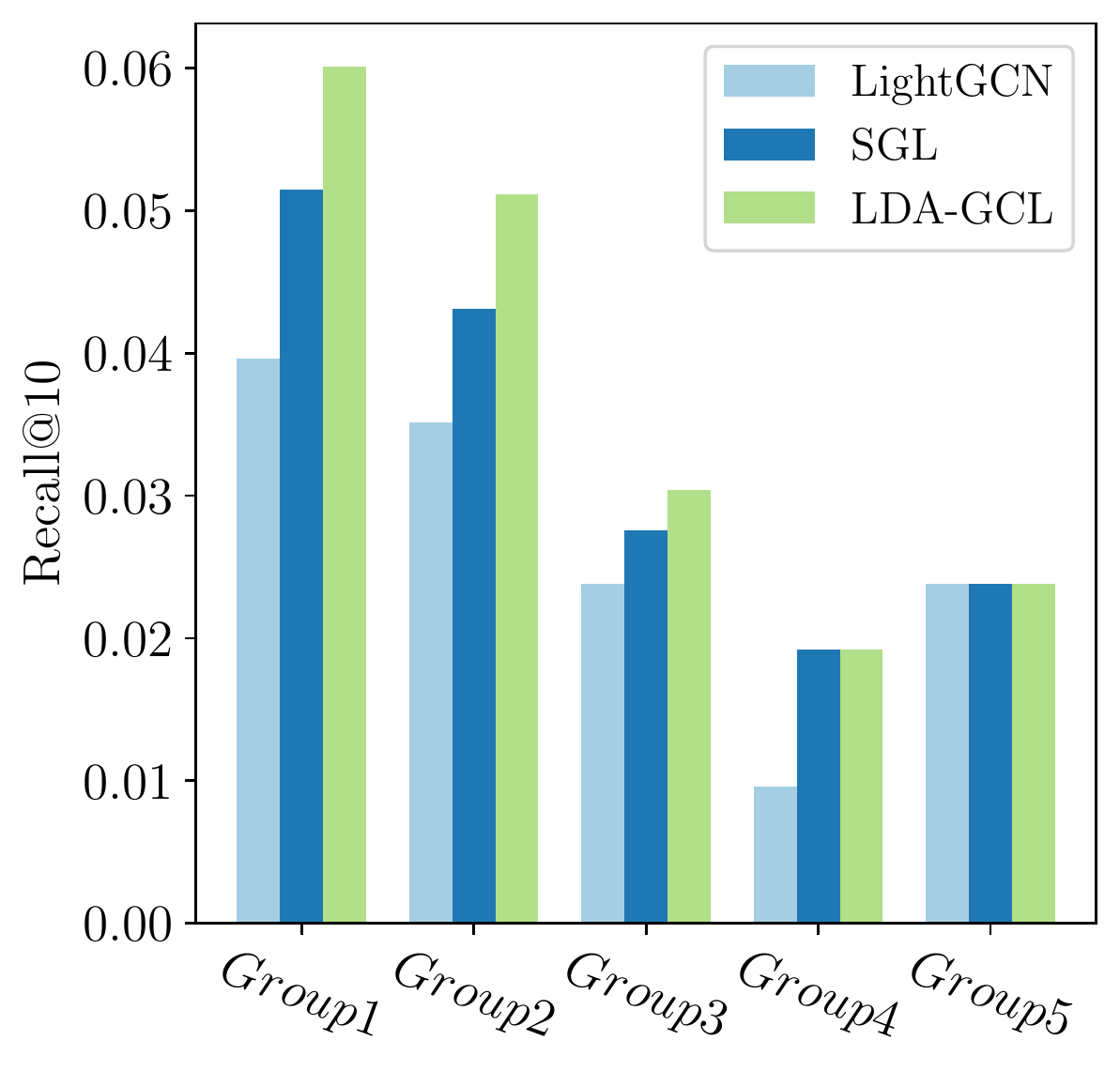}
  \caption{\hspace{-0.7em} Alibaba-iFashion}
  \label{fig:group-analysis-2}
  \end{subfigure}
  \caption{Performance analysis over different users groups.
  $G_1$ is the group of users with the \textit{lowest} interaction number.}
  \label{fig:group-analysis}
\end{minipage}
  \hfill
\begin{minipage}[t]{0.48\linewidth}
\centering
    \begin{subfigure}[t]{0.48\linewidth}
    \centering
    \includegraphics[width=\linewidth]{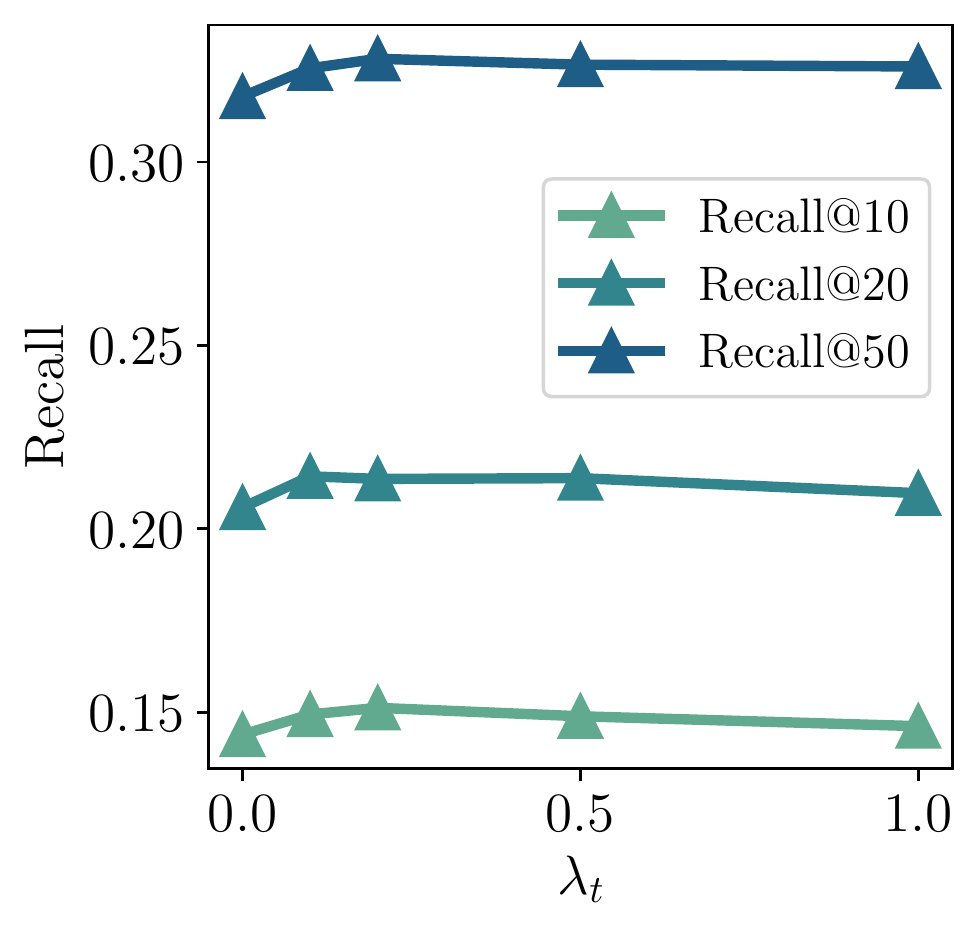}
    \caption{Gowalla}
    \label{fig:layer}
  \end{subfigure}
  \hfill
  \begin{subfigure}[t]{0.48\linewidth}
    \centering
    \includegraphics[width=\linewidth]{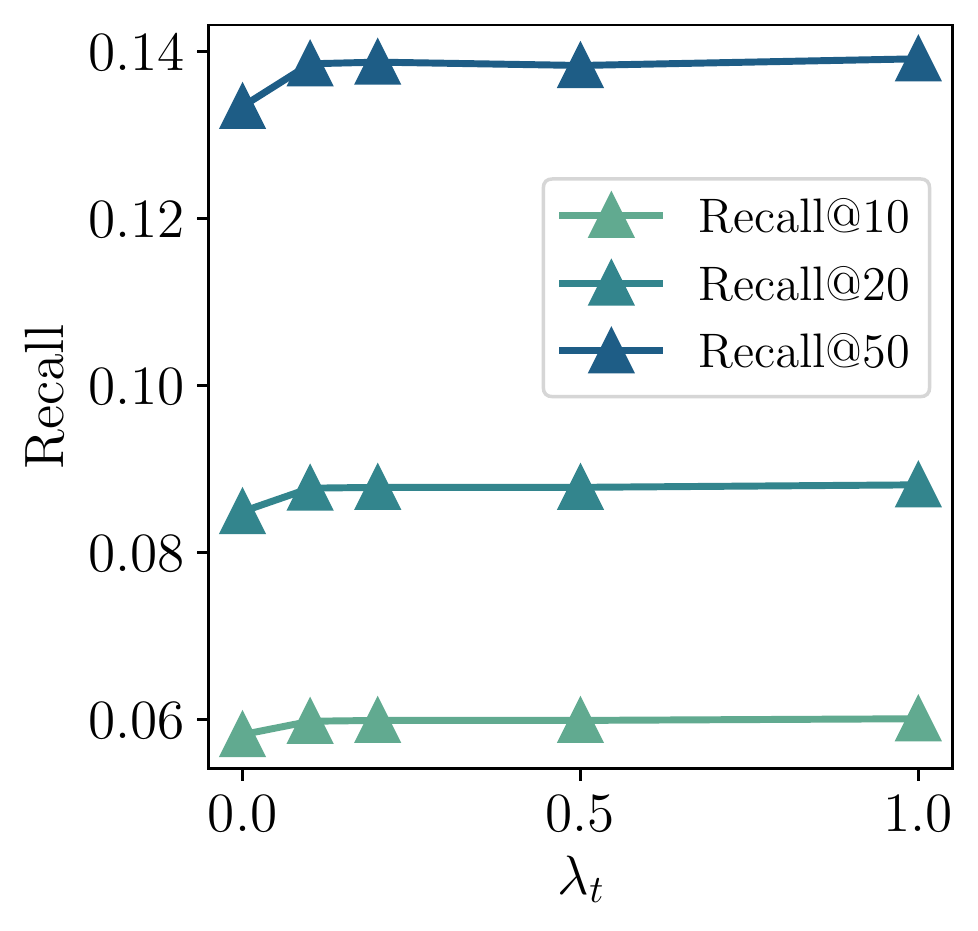}
    \caption{\hspace{-0.7em} Alibaba-iFashion}
    \label{fig:d}
  \end{subfigure}
  \caption{Parameter Analysis of $\lambda_t$.}
  \label{fig:lambda_analysis}
\end{minipage}
\vspace{-1em}
 \end{figure}

In general, graph contrastive learning on recommender systems can alleviate the problem of data sparseness commonly found in recommender systems~\cite{wu2021self-sigir,lin2022improving}.
To further verify the proposed \method~ can alleviate the sparsity of interaction data, we evaluate the performance of the different groups of users. 
We split all the users into five groups based on their interaction numbers.
Then, we compare the performance of LightGCN, SGL, and \method~ on these groups.
From \figref{fig:group-analysis}, we can find that the performance improvements of SGL mainly come from recommending the items with sparse interactions, which is consistent with findings in ~\cite{lin2022improving}.
On the Gowalla dataset, the performance of SGL is not even as good as LightGCN on the groups of users with a high number of interactions (\eg $Group3, Group4$).
For our \method, we can find the performance of \method~ is consistently better than or comparable with LightGCN and SGL on all user groups.
Besides, as the number of interactions decreases, \method~ achieves greater improvements.
For instance, on the Alibaba-iFashion dataset, \method~ outperform LightGCN 27.9\%, 45.5\%, and 51.8\%  on $Group3$, $Group2$, and $Group1$, respectively.
In conclusion, \method~ can further alleviate the problem of data sparsity compared to SGL.

\subsubsection{Ablation Study}

\label{sec:ablation_study}
\vspace{-1em}
As we discussed before, \method~consists of two parts: learning data augmentation and graph contrastive learning. 
We do ablation studies to investigate the role of each part.

\begin{table}
\vspace{-1em}
\centering
\caption{Performance comparison of different variants of \method.}
\label{tab:ablation}
\scalebox{0.85}{
\begin{tabular}{c|cc|cc}
\toprule
\multicolumn{1}{c}{\multirow{2}{*}{Method}} & \multicolumn{2}{c}{Gowalla}                            & \multicolumn{2}{c}{Alibaba-iFashion}                                    \\
\multicolumn{1}{c}{}                        & \multicolumn{1}{c}{Recall@10} & \multicolumn{1}{c}{NDCG@10} & \multicolumn{1}{c}{Recall@10} & \multicolumn{1}{c}{NDCG@10} \\ \midrule
LightGCN & 0.1342	& 0.0962 & 0.0395	& 0.0212 \\
\midrule
DA-GCL(0.0,0.0) & 0.1488 & 0.1085 & 0.0497 & 0.0274 \\
DA-GCL(0.1,0.0) & 0.1492 & 0.1083 & 0.0529 & 0.0289 \\
DA-GCL(0.0,0.1) & 0.1487 & 0.1067 & 0.0544 & 0.0299 \\
DA-GCL(0.1,0.1) & 0.1479 & 0.1063 & 0.0553 & 0.0303 \\
DA-GCL(0.0,0.5) & 0.1412 & 0.1010 & 0.0533 & 0.0290 \\
DA-GCL(0.1,0.5) & 0.1409 & 0.1003 & 0.0542 & 0.0296 \\
DA-GCL(0.0,1.0) & 0.1369 & 0.0973 & 0.0520 & 0.0282 \\
DA-GCL(0.1,1.0) & 0.1359 & 0.0963 & 0.0526 & 0.0285 \\
\midrule
LDA-GCL (w NGCF) & 0.1488 & 0.1078 & 0.0589 & 0.0322 \\
LDA-GCL (w/o EA) & 0.1499 & 0.1087 & 0.0579 & 0.0319 \\
LDA-GCL & 0.1512 & 0.1090 & 0.0599 & 0.0330 \\
\bottomrule
\end{tabular}
}
\end{table}
First, since our edge candidates include original edges and added edges, we can remove the added edges from edge candidates.
In other words, we use only edge-dropping data augmentation. 
We mark it as LDA-GCL (w/o EA).
Second, we use NGCF as the pre-trained GNN to generate edge candidates, which is marked as LDA-GCL (w NGCF).
Third, we can use randomly added and dropped edges for data augmentation and remove the module for learning data augmentation.
We mark such data augmentation methods as DA-GCL($p_A$, $p_D$), where $p_A$ is the ratio of added edges, and $p_D$ is the ratio of dropped edges.
DA-GCL is close to SGL, the difference is that SGL only includes the edge-dropping data augmentation, which happens on two views.
In contrast, DA-GCL has both edge-dropping and edge-adding on only one view.
The other view in DA-GCL is the original graph.
We vary $p_D$ from \{0.0, 0.1\}\footnote{Large $p_D$ will harm the performance according to the report in \cite{wu2021self-sigir}} and vary $p_A$ from \{0.0, 0.1, 0.5, 1\} . 
Note that, $p_A=0.0$ and $p_D=0.0$ mean that no data augmentation is employed, but the contrastive loss is still employed.

The results are reported in \tableref{tab:ablation}.
From \tableref{tab:ablation}, we can find that: 
\begin{enumerate*}[label=(\arabic*)]
    \item All the variants of~\method~outperform LightGCN by large margins, which demonstrates the effectiveness of contrastive learning in improving  performance.
    \item When we do not adopt the edge-adding strategy, the performance of LDA-GCL (w/o EA) shows a degradation compared to LDA-GCL.
    On Recall@10, LDA-GCL outperforms LDA-GCL (w/o EA) by 0.87\% and 	3.5\% on the dataset Gowalla and Alibaba-iFashion, respectively. 
    It demonstrates that the strategy of adding edges achieves greater improvement on more sparse datasets.
    \item Compared with LDA-GCL, there is a drop in performance for LDA-GCL (w NGCF). 
    It shows that better edge candidates have an impact on the results.
    \item We can find that learnable data augmentation  methods (\ie \method) show better performance than randomly sampling data augmentation methods (\ie DA-GCL) on both datasets.
    \item When we remove the learning data augmentation module, we can find that the edge-adding strategy (\eg $p_A=0.1$) performs better on the sparser Alibaba-iFashion dataset compared to the edge-dropping strategy (\eg $p_A=0.0$).
    Performance decreases as the ratio of added edges increases. 
    We speculate that this is due to the introduction of more noise.
    \item Surprisingly, when the graph augmentation is detached (\ie DA-GCL(0.0, 0.0)), the performance gains are still so remarkable. 
    This result was also reported in~\cite{yu2022graph}, which suggests the potential of a proper graph augmentation.
\end{enumerate*}

\subsection{Parameter Analysis (RQ3)}

We analyze the  hyper-parameters of $\lambda_t$ in \equref{eq:loss_function}.
We vary it from \{0.0, 0.1, 0.2, 0.5, 1.0\}.
Note that,  $\lambda_t = 0.0$ means there is no $\min_{t}\ I(f(G); f(t(G)))$.
Due to the page limits, we report the Recall in \figref{fig:lambda_analysis}. 
The NDCG metrics have similar findings.

In \figref{fig:lambda_analysis}, the best settings of $\lambda_t$ on the Alibaba-iFashion dataset and the Gowalla dataset are 1.0 and 0.2, respectively.
When we adopt $\min_{t}\ I(f(G); f(t(G)))$ (\ie $\lambda_t > 0.0$), the hyper-parameter $\lambda_t$ makes less of a difference.
We can find that the performance is worst when $\lambda_t=0.0$ on both datasets.
More specifically, the settings of $\lambda_t=0.1$ outperform the settings of $\lambda_t=0.0$, 2.75\%, 3.30 \% and 3.82\% on Recall@10, Recall@20, and Recall@50 on the Alibaba-iFashion dataset.
The settings of $\lambda_t=0.1$ outperform the settings of $\lambda_t=0.0$, 3.82\%, 4.13 \% and 2.42\% on Recall@10, Recall@20, and Recall@50 on the Gowalla dataset.
It  demonstrates the effectiveness of LDA-GCL adopting the InfoMin principle guided data augmentation learning.

\section{Related Work}
\label{sec:related_work}

\subsection{GNN-based  Recommendation}
Nowadays, GNNs are also widely used in recommender systems.
Different from traditional CF methods, such as matrix factorization (MF) methods~\cite{rendle2012bpr,koren2009matrix,he2017neural} and auto-encoder (AE) methods~\cite{liang2018variational}, Graph Neural Networks (GNN) are used to model interaction data into a bipartite graph and learn  users and items effective representations from the graph structure information~\cite{wang2019neural,he2020lightgcn,wang2020disentangled}.
Most GNN methods in recommender system follow the message-passing scheme~\cite{gilmer2017neural} to utilize the bipartite graph structure, including the design of aggregate functions and update functions for recommendations.
Most representatively, LightGCN~\cite{he2020lightgcn} removes the self-connection, feature transformation, and nonlinear activation of standard GCNs, and leverages the high-order graph structure to enhance the recommendation performance.
LightGCN has also been used as a backbone model for many further works (\eg GCL~\cite{wu2021self-sigir}).
Although the GNN models are effective, they still suffer from data sparsity issues in recommender systems.

\subsection{Contrastive Learning in Recommendation}
Contrastive Learning (CL) as a self-supervised manner~\cite{yu2022self}, has been applied in Recommender Systems (RS)~\cite{yao2021self,yu2021self,lin2022improving}.
In recommender system scenarios, Graph Contrastive Learning (GCL) is often used to alleviate the data sparsity problem~\cite{lin2022improving,wu2021self-sigir}.
For example, SGL~\cite{wu2021self-sigir} constructs the contrastive pairs by random sampling.
The methods aforementioned for data augmentation are manually designed and require expertized knowledge.
And improper data augmentation will hinder the performance of GCL~\cite{yu2022graph}.
The proper way of data augmentation requires guiding principles instead of the heuristic design.

Inspired by \textbf{InfoMin} principle proposed by~\cite{tian2020makes}, AD-GCL~\cite{suresh2021adversarial} optimizes adversarial graph augmentation strategies to train GNNs to avoid capturing redundant information during the training.
However, AD-GCL is designed to work on unsupervised graph classification with lots of small graphs, under the pre-training \& fine-tuning scheme.
But in graph collaborative filtering in recommendation (link prediction with a large graph, under the joint learning scheme), how to design a learnable data augmentation method for better performance improvement is an open and promising research problem.

\section{Conclusion and Future Work}
\label{sec:conclusion}

In this work, we develop a theoretically motivated learnable data augmentation model for GCL in recommendation, instead of  heuristic designs.
Guided by both  \textbf{InfoMin} and \textbf{InfoMax} principles, our model is an adversarial framework that can better enhance the effect of GCL in the recommendation.
Via learning better data augmentation in GCL, our model achieves state-of-the-art performance on several public benchmark datasets.
Further experiments on each component demonstrate the effective design of \method.
Our methods open the door to designing learnable data augmentation methods instead of heuristic augmentation methods in recommendation.

The main limitation of our model is the high complexity of learning data augmentation, which may cause low training efficiency, compared with simple data augmentation~\cite{yu2022graph}.
We can make improvements on the efficiency in future work.
A potential boosting scheme is the pre-trained edge operator models.

\section*{Acknowledgement}
This work is funded by the National Natural Science Foundation of China under Grant Nos. 62102402, U21B2046, 62272125, and the National Key R\&D Program of China (2022YFB3103701). 
Huawei Shen is supported by Beijing Academy of Artificial Intelligence (BAAI).
Junjie Huang is supported by Tencent Rhino-Bird Elite Training Program.

\bibliographystyle{splncs04}
\bibliography{refs}

\end{document}